\documentclass[nofootinbib,twocolumn,10ppt]{revtex4}%
\usepackage{amsfonts}
\usepackage{amsmath}
\usepackage{amssymb}
\usepackage{graphicx}%
\providecommand{\U}[1]{\protect\rule{.1in}{.1in}}
\begin{document}
\title{Thermodynamics of a Photon Gas in Nonlinear Electrodynamics}
\author{P. Niau Akmansoy}
\email{pniau7@gmail.com}
\affiliation{Departamento de F\'{\i}sica Te\'{o}rica e Experimental, Universidade Federal
do Rio Grande do Norte. Campus Universit\'{a}rio s/n, CEP 59072-970, Natal, Brazil}
\author{L. G. Medeiros}
\email{leogmedeiros@ect.ufrn.br}
\affiliation{Escola de Ci\^{e}ncia e Tecnologia, Universidade Federal do Rio Grande do
Norte. Campus Universit\'{a}rio s/n, CEP 59072-970, Natal, Brazil}

\pacs{41.20.Jb, 64.60.De, 42.25.Lc}

\begin{abstract}
In this paper we analyze the thermodynamic properties of a photon gas under
the influence of a background electromagnetic field in the context of any
nonlinear electrodynamics. Neglecting the self-interaction of photons, we
obtain a general expression for the grand canonical potential. Particularizing
for the case when the background field is uniform, we determine the pressure
and the energy density for the photon gas. Although the pressure and the
energy density change when compared with the standard case, the relationship
between them remains unaltered, namely $\rho=3p$. Finally, we apply the
developed formulation to the cases of Heisenberg-Euler and Born-Infeld nonlinear
electrodynamics. For the Heisenberg-Euler case, we show that our formalism
recover the results obtained with the $2$-loop thermal effective action approach.
\end{abstract}
\endpage{ }
\maketitle

\section{Introduction}

Maxwell electrodynamics is one of the most successful theories in the history
of physics. Classically, it is able to describe all the known electric and magnetic
phenomena including the creation and propagation of electromagnetic
wave. Its quantum version, QED, is the most successful QFT ever built and
tested producing accurate results up to ten parts in a billion \cite{PDG}.
Nevertheless, alternatives and extensions of Maxwell electrodynamics have been
proposed since its creation in the second half of the nineteenth century. The
motivations for proposing these extensions are quite diverse and include the
problem of divergence for the classical Coulomb potential
\cite{ClaDiv1,BI,BI1}, experimental constrains for the photon mass
\cite{PRL,RevPhoMass,ProPod1,ProPod2}, the classic study of vacuum
polarization effects \cite{Schwinger,Dune2004,Holger book}, modifications on
electrodynamic in the context of branes \cite{Zwieb}, etc.

From the point of view of a gauge theory \cite{Utiyama}, Maxwell
electrodynamics arises imposing four conditions to its Lagrangian: \emph{(i)}
the Lagrangian $L$ must be Lorentz invariant; \emph{(ii)} $L$ should be gauge
invariant for $U(1)$ symmetry group; \emph{(iii)} $L$ must depend only on
$A_{\mu}$ and its first derivative; \emph{(iv)} the Lagrangian should contain
only quadratic forms of $A_{\mu}$ and its first derivative. There are many
examples of extensions of Maxwell electrodynamics which break, at least, one
of the above four conditions. Violation of the first condition is usually done
introducing tensor quantities that create preferred directions and
boost-dependent effects \cite{Kost,VL1,VL2,VL3}. On the other hand, if it is
introduced an usual term of mass for the photon - Proca electrodynamics
\cite{Proca} - the second condition is necessarily violated. Moreover, if only
the third condition is broken a Maxwell extension emerges known as Podolsky
electrodynamics \cite{Pod1,Pod2,Pod3}. Finally, if only the fourth condition
is violated then a class of electrodynamics called nonlinear electrodynamics
(NLED) arises \cite{Plebansky}. An example of NLED is Born-Infeld theory
\cite{BI,BI1}. This paper has NLED as background.

Models involving NLED appear in different branches of physics, but the most
important situation in which they arise is in the context of vacuum polarization.
Since the early thirties, we know that virtual electron loops induce a
self-coupling of the electromagnetic field. In the energy scales below the
electron mass and for a constant electromagnetic field this self-interaction
could be represented by an effective field theory known as Heisenberg-Euler
NLED. The one and two loops Heisenberg-Euler effective actions were calculated
in \cite{Euler-Heisenberg,Schwinger,Weis} and
\cite{Ritus,Dittrich,Re2Sch,Kors} respectively. Besides vacuum polarization in
QED, models of NLED emerge in other contexts such as string theories
\cite{FraTsey,Abjorn,SiWi,Tsey99} and in the description of radiation
propagation inside specific materials \cite{Lor,LoKli1,NovBi,NoSalBerLoKli}.

Recently, in the context of deformed special relativity (DSR), some papers
have studied thermodynamics consequences of modifications in the relativistic
dispersion relation \cite{Ko,CaMa,DSR PhoGas,ChaChat}. These works are usually
motivated by the possibility of Lorentz symmetry breakdown at Planck scales
\cite{Cam1,Cam2}. Modified dispersion relations appear not only in DSR models
but also in NLED scenarios. This interesting effect occurs in NLED when we
study wave propagation in an electromagnetic background. Considering a background
electromagnetic field it is possible to show that the interaction between a
electromagnetic wave and this field produces a modified dispersion relation
\cite{Boilat,Birula,Novello2000}.

In the present work, we explore the thermodynamics properties of a photon gas
under the influence of a background electromagnetic field in NLED scenarios. The
paper is organized as follows. In section $2$, we present the modified
dispersion relations and perform the thermodynamic analysis for a photon gas
in an arbitrary NLED background. In section $3$, we apply the developed
formulation on Heisenberg-Euler NLED and compare the results with the
effective field theory at finite temperature. Section $4$ presents a
comparison among the thermodynamic properties of the photon gas in the
background of Maxwell and Born-Infeld electrodynamics. The final
remarks and further perspectives are made in section $5$. Finally, a
discussion on the birefringence phenomenon is presented in Appendix $A$.

\section{Photon Gas Thermodynamics}

A general NLED\ is described in terms of the Lagrangian $L\left(  F,G\right)
$ where $F=-\frac{1}{4}f^{\mu\nu}f_{\mu\nu}=\frac{1}{2}\left(  E^{2}%
-B^{2}\right)  $ and $G=-\frac{1}{4}\hat{f}^{\mu\nu}f_{\mu\nu}=\vec{E}%
\cdot\vec{B}$ are the contractions between the electromagnetic field tensor
and its dual. The first step to obtain the modified dispersion relation is to
introduce an energy scale $M$ and require that all other energy scales of
the system (with exception of the field strength) are small when compared with $M$.\footnote
{This energy scale is specified in each NLED.} This condition ensures that the electromagnetic field is slowly
varying and restricts the radiation to be low frequency when compared with $M$ \cite{Holger book}.
The next step is split the field strength $f^{\mu\nu}$ into a background field
$F^{\mu\nu}$ and a plane wave $\phi^{\mu\nu}$. Finally, considering a linear
approximation in $\phi^{\mu\nu}$ we obtain%

\begin{equation}
\left[  \eta^{\mu\nu}+z_{\pm}\left(  F,G\right)  F^{\mu\alpha}F_{\alpha
}^{\text{ }\nu}\right]  k_{\mu}k_{\nu}=0, \label{Metrica Efetiva}%
\end{equation}
where
\begin{equation}
z_{\pm}\left(  F,G\right)  \equiv\frac{-2F\sigma+L_{F}\left(  L_{GG}%
+L_{FF}\right)  \pm\sqrt{\delta}}{2\left[  L_{F}^{2}+2L_{F}\left(
L_{FG}G-L_{GG}F\right)  -G^{2}\sigma\right]  } \label{z main}%
\end{equation}
with $L_{F}=\frac{\partial L}{\partial F}$, $L_{G}=\frac{\partial L}{\partial
G}$,%
\begin{equation}
\delta=\left[  L_{F}\left(  L_{FF}-L_{GG}\right)  -2F\sigma\right]
^{2}+4\left[  L_{F}L_{FG}-G\sigma\right]  ^{2}\text{,}\nonumber
\end{equation}
and%
\[
\sigma=L_{FF}L_{GG}-L_{FG}^{2}.
\]
This result was first obtained by the authors in \cite{Birula}.

The function $z_{\pm}\left(  F,G\right)  $ contains all the information of the
NLED and the $\pm$ sign indicates the existence of two possible solutions. Besides,
it is interesting to note that the linear approximation does not imply that
the $F^{\mu\nu}$ is much stronger than the $\phi^{\mu\nu}$. Thus, the zero-field
limit for the background field is well defined.

Using Minkowski's signature, $\eta^{\mu\nu}=diag\left(  +,-,-,-\right)  $, and
the 4-wave vector $k^{\mu}=\left(  \omega,\vec{k}\right)  $,
(\ref{Metrica Efetiva}) can be rewritten to obtain the energy $\omega$ of a
photon in terms of $z_{\pm}\left(  F,G\right)  $, the wave vector of the
photon, and the background electromagnetic fields, $\vec{E}$ and $\vec{B}$
as\footnote{The units used in this paper are $c=\hslash=1$.}
\begin{equation}
\omega_{\pm}=\frac{z_{\pm}S+\sqrt{z_{\pm}^{2}S^{2}-\left[  1+\vec{E}^{2}%
z_{\pm}\right]  \left\{  z_{\pm}R-\vec{k}^{2}\right\}  }}{1+\vec{E}^{2}z_{\pm
}} \label{omega}%
\end{equation}
where%
\[
S=\left(  \vec{E}\times\vec{B}\right)  \cdot\vec{k},\text{ \ }R=\left(
\vec{k}\times\vec{B}\right)  ^{2}-\left(  \vec{k}\cdot\vec{E}\right)  ^{2}.
\]

The expression (\ref{omega}) deserves some comments. First of all, there is
not only one dispersion relation but two of them. It occurs because each
transverse mode of wave propagation has a different dispersion relation,
\textit{i.e.} the energy $\omega$ depends on the photon polarization. This
phenomenon is called birefringence and it is present in NLED models in general
\cite{CLP} (for details see appendix \ref{apen}). Moreover, these dispersion
relations are coordinate dependents in general. The photon energy depends on
the points of space-time because of its interaction with the background field.
This is anticipated since the self-interaction processes of electromagnetic field
are present in any nonlinear models.

\subsection*{Statistical Approach}

With the general form of the photon dispersion relation in terms of NLED
properties and the background field we now must resort to a statistical
approach in order to obtain the thermodynamic properties of the photon gas.
Since the bosonic nature of the photon is not affected by the generalization
of the underlying dynamics, the Bose-Einstein statistics is still
appropriate. There is an important point concerning the choice of reference
frame. Although the phase volume $d^{3}xd^{3}p$ is Lorentz-invariant
\cite{MisThor} the same can not be said about the dispersion relation
(\ref{omega}). Indeed, the dependence of $\omega_{\pm}$ on the background
electromagnetic field makes it clearly dependent on the reference frame. In
order to describe the thermal radiation, we must choose a reference frame
comoving with the matter which produced the gas of photons. Thus the photons
with the same energy are \emph{isotropically} distributed in the space. It
defines the reference frame in which the partition function must be calculated.
Note that the same choice is done when we perform the usual calculation in
Maxwell theory.\footnote{As it can be verified in the measurement of CMB, the
effect of a different choice of reference frame distorts the black body radiation
distribution.}.

In the grand canonical ensemble \cite{Pat}, the potential for the gas of
photons is given by%
\[
\Omega=-\frac{g}{V}{\sum\limits_{j}}\ln\left(  1-e^{-\beta\omega_{j}}\right)
\]
where $\beta=\frac{1}{kT}$, $V$ is the volume and $g$ takes account the
internal degrees of freedom. In NLED each transverse mode corresponds to a
different dispersion relation, so $g=1$ and the expression above can be
rewritten as
\begin{align}
\Omega &  =\frac{-1}{\left(  2\pi\right)  ^{3}V}\left[  \int\ln\left(
1-e^{-\beta\omega_{+}}\right)  d^{3}xd^{3}p\right.  +\nonumber\\
&  +\left.  \int\ln\left(  1-e^{-\beta\omega_{-}}\right)  d^{3}xd^{3}p\right]
, \label{macro}%
\end{align}
It is clearly seen that in the cases of non-birefringent NLED, the grand
canonical potential falls back to the known expression. The thermodynamical
quantities are computed in the usual way i.e., the pressure $p$ and the energy
density $\rho$ are given by%
\begin{equation}
p=\frac{\Omega}{\beta}\text{ \ \ and \ \ }\rho=-\left(  \frac{\partial\Omega
}{\partial\beta}\right)  . \label{def pre den ener}%
\end{equation}

\subsubsection*{Uniform electromagnetic field case}

Although the equation (\ref{macro}) represents the general case, it is
difficult to calculate the integrals mainly due to the spatial dependence of
the electromagnetic field. Thus, for simplification purposes we consider a
particular case where the electromagnetic field is uniform. Without loss of
generality, it is possible to align the $x$ axis with the electric field and
the magnetic field within the $xy$ plane. Then, equation
(\ref{omega}) reduces to

\[
\omega_{\pm}=\frac{y_{\pm}k_{3}+\sqrt{a_{\pm}k_{1}^{2}+b_{\pm}k_{2}^{2}%
+c_{\pm}k_{3}^{2}+e_{\pm}k_{1}k_{2}}}{x_{\pm}},
\]
where the subscript indicates the cartesian components and%

\begin{align*}
a_{\pm}  &  =\left(  1+E_{1}^{2}z_{\pm}-B_{2}^{2}z_{\pm}\right)  x_{\pm
}\text{,\ \ \ \,}e_{\pm}=2x_{\pm}B_{1}B_{2}z_{\pm}\text{,}\\
b_{\pm}  &  =\left(  1-B_{1}^{2}z_{\pm}\right)  x_{\pm}%
\text{,\ \ \ \ \ \ \ \ \ \ \ \ \ \,\,}x_{\pm}=1+E_{1}^{2}z_{\pm}\text{,}\\
c_{\pm}  &  =x_{\pm}-x_{\pm}B_{1}^{2}z_{\pm}-B_{2}^{2}z_{\pm}%
\text{,\ \ \ \ \,}y_{\pm}=E_{1}B_{2}z_{\pm}\text{.}%
\end{align*}
Note that all these factors are constants.

Due the similarity of the terms in the r.h.s. of (\ref{macro}), one can write
\begin{equation}
\Omega_{\pm}=-\frac{1}{\left(  2\pi\right)  ^{3}}\int\ln\left(  1-e^{-\beta
\omega_{\pm}}\right)  d^{3}k \label{Aux1}%
\end{equation}
with%
\[
\Omega=\Omega_{+}+\Omega_{-}.
\]
Expanding the $\ln(...)$ in Taylor series and performing some variable
substitutions the integral (\ref{Aux1}) is rewritten as
\[
\Omega_{\pm}=A\sum\limits_{n=1}^{\infty}\frac{1}{n}\left[  {\int
\limits_{0}^{\pi}}{\int\limits_{0}^{\infty}}e^{-\frac{n\beta}{x_{_{\pm}}%
}\left(  1+\frac{y_{\pm}}{\sqrt{c_{\pm}}}\cos\theta\right)  r}r^{2}\sin\theta
drd\theta\right]  ,
\]
where the constant $A$\ is

\[
A=\frac{1}{4\pi^{2}\sqrt{c_{\pm}\left(  4a_{\pm}b_{\pm}-e_{\pm}^{2}\right)  }%
}.
\]
This integral converges under the requirements
\begin{equation}
x_{\pm}>0\text{ \ and \ }\left\vert y_{\pm}\right\vert <\sqrt{c_{\pm}}.
\label{Cons}%
\end{equation}
The result is:

\begin{equation}
\Omega_{\pm}=\frac{\pi^{2}}{90}\frac{\left\vert c_{\pm}\right\vert }{\left(
1-z_{\pm}\left(  B_{1}^{2}+B_{2}^{2}\right)  \right)  ^{2}}\frac{1}{\beta^{3}%
}. \label{Omega}
\end{equation}

The potential above was calculated on the black body rest frame. On the other
hand, it is possible to choose a new frame which has a $4$-velocity $u^{\mu}$
(in the black body rest frame $u^{\mu}=\left(  1,0,0,0\right)  $). Thus, with
$u^{\mu}$ we can define a new invariant
\[
H=\left(  u_{\mu}F^{\mu\alpha}\right)  \left(  u_{\nu}F_{\text{ \ }\alpha
}^{\nu}\right)
\]
and rewrite (\ref{Omega}) in terms of invariants $F$, $G$ and
$H$:%
\begin{equation}
\Omega_{\pm}=\frac{\pi^{2}}{90}\frac{\left\vert 1+2z_{\pm}F-z_{\pm}^{2}%
G^{2}\right\vert }{\left(  1+z_{\pm}\left(  2F-H\right)  \right)  ^{2}}%
\frac{1}{\beta^{3}}\equiv\frac{K_{\pm}}{\beta^{3}}. \label{K definition}%
\end{equation}
Because of the dependence of the constraints on $z_{\pm}$, they must be
analyzed separately for each NLED. Two examples are given in the next sections.

Substituting the explicit form of $\Omega$ in (\ref{def pre den ener}) it is
possible to obtain the pressure,%
\begin{equation}
p\left(  kT\right)  =\left(  K_{+}+K_{-}\right)  \left(  kT\right)  ^{4},
\label{pressure}%
\end{equation}
and the energy density%
\begin{equation}
\rho\left(  kT\right)  =3\left(  K_{+}+K_{-}\right)  \left(  kT\right)  ^{4}.
\label{energy density}%
\end{equation}
If we set $z_{\pm}=0$ it follows $K_{+}=K_{-}=\frac{\pi^{2}}{90}$ and the
usual results are recovered.

An interesting result is that, despite of $p$ and $\rho$ being affected by the
constant background field, the equation of state remains unaffected, i.e.
\[
p=\frac{\rho}{3}.
\]
This result is independent from a particular NLED.

Before the end of this section, it is interesting to reanalyze the approximation
made at the beginning. To obtain the dispersion relation (\ref{Metrica Efetiva})
it was supposed that the frequency associated with the radiation is smaller than the
energy scale $M$. It implies that the approach developed is only valid when the
average energy per photon $\varepsilon$
is much smaller than $M$. Thus, remembering that the photon numerical density $n$
is proportional to $\left(  kT\right)  ^{3}$ we establish the following condition:
\begin{equation}
\varepsilon \sim\frac{\rho}{n} \sim\left(  kT\right) \ll M.\label{C1}%
\end{equation}

\section{Heisenberg-Euler Effective Lagrangians}

In this section, we apply the above formulation to Heisenberg-Euler NLED.
The $1$-loop Heisenberg-Euler effective action \cite{Schwinger,Dune2004} is
given by%
\begin{align}
L^{1}  & =F+\frac{e^{2}ab}{8\pi^{2}}{\int\limits_{0}^{\infty}}\frac{ds}%
{s}e^{-im^{2}s}\nonumber\\
& \times \left[  \cot\left(  eas\right)  \coth\left(  ebs\right)  -\frac{1}%
{ab}\left(  \frac{1}{e^{2}}+\frac{2}{3}F\right)  \right]  \label{L1}%
\end{align}
where%
\begin{equation}
a^{2}=\sqrt{F^{2}+G^{2}}-F\text{, \ }b^{2}=\sqrt{F^{2}+G^{2}}+F.\nonumber
\end{equation}
The constants $e$ and $m$ are respectively the charge and mass of the electron.
It is worth noting that due to the fact $G$ gains a minus sign after a parity
transformation, the theory must only depend on $G^2$ to be parity invariant.

It is important to emphasize that the Heisenberg-Euler NLED is an effective
description of QED in the low-energy regime. More precisely, the Lagrangian
(\ref{L1}) is only valid for photons with energy lower than the electron mass. This
characteristic scale sets $M=m$, and thus our approach is valid only in the
regime of $kT \ll m$.

Expanding (\ref{L1}) in power series and integrating over $s$ leads to
\begin{equation}
L^{1}=F+c_{1}F^{2}+c_{2}G^{2}+c_{3}F^{3}+c_{4}FG^{2}+... \label{L1series}%
\end{equation}
with%
\[
c_{1}=\frac{8\alpha^{2}}{45m^{4}}\text{,\ }c_{2}=\frac{14\alpha^{2}}{45m^{4}%
}\text{, }c_{3}=\frac{2^{8}\pi\alpha^{3}}{315m^{8}}\text{,\ }c_{4}=\frac
{2^{5}\times13\pi\alpha^{3}}{315m^{8}}.
\]
where $\alpha$ is the structure constant of QED.

Let us concentrate on a weak-field analysis expanding $K_{\pm}$ to
quadratic terms in the invariants:
\begin{align}
K_{\pm}  & \simeq\frac{\pi^{2}}{90}\left[  1+2z_{\pm}\left(  H-F\right)
\right.  \nonumber\\
& +\left.  z_{\pm}^{2}\left[  3\left(  H-2F\right)  ^{2}-G^{2}+4\left(
H-2F\right)  F\right]  \right]  +\mathcal{O}(3).\label{Kappro}%
\end{align}
By $\mathcal{O}(3)$ we mean cubic combinations of invariants $F$, $G$ and $H$.
The coefficient $z_{\pm}$ should be linear in the invariants which means that
\begin{align*}
z_{+}  &  =2c_{1}+2\left(  2c_{2}^{2}-4c_{1}c_{2}+3c_{3}\right)
F+\mathcal{O}(2)\text{,}\\
z_{-}  &  =2c_{2}+2\left(  2c_{2}^{2}-2c_{1}^{2}+2c_{1}c_{2}+c_{4}\right)
F+\mathcal{O}(2).
\end{align*}
Besides, as we are concerned with first corrections to Maxwell electrodynamics
all terms greater than $\alpha^{3}$ will be neglected. Thus, (\ref{Kappro}) is
rewritten as%
\begin{align*}
K_{+}  &  \simeq\frac{\pi^{2}}{90}\left[  1+4c_{1}\left(  H-F\right)
+12c_{3}\left(  H-F\right)  F+\mathcal{O}(3)\right] \\
K_{-}  &  \simeq\frac{\pi^{2}}{90}\left[  1+4c_{2}\left(  H-F\right)
+4c_{4}\left(  H-F\right)  F+\mathcal{O}(3)\right]
\end{align*}
and (\ref{energy density}) leads to%
\begin{align}
\rho\left(  kT\right)   &  =\frac{\pi^{2}}{15}\left(  kT\right)  ^{4}%
+\frac{44\alpha^{2}\pi^{2}}{675}\frac{\left(  H-F\right)  }{m^{4}}\left(
kT\right)  ^{4}\nonumber\\
&  +\frac{2^{6}\times37\alpha^{3}\pi^{3}}{3^{3}\times5^{2}\times7}%
\frac{F\left(  H-F\right)  }{m^{8}}\left(  kT\right)  ^{4}+\mathcal{O}(3).
\label{rhoHES}%
\end{align}
Note that $\left(  H-F\right)  $ is always positive.

This result is in complete agreement with the calculation of QED effective
action at finite temperature. In fact, it has been shown in \cite{Gies1999}
that in the thermal $1$-loop QED effective Lagrangian at low-temperature expansion
all the terms are damped by a factor $e^{-\frac{m}{kT}}$. However,
the thermal $2$-loop effective action at low-temperature
expansion in the weak-field limit produces a dominant term exactly as
presented in (\ref{rhoHES}).\footnote{Equation $(32)$ in \cite{Gies2000} with
$\rho=T\left(  \frac{\partial L^{2T}}{\partial T}\right)  -L^{2T}.$} We do not obtain terms proportional to $\left(
kT\right)  ^{6}$ presented in the thermal $2$-loop effective action\footnote{Second terms at low-temperature
expansion - Equation $(33)$ in \cite{Gies2000}.} because our approach is valid only in
the regime of $kT \ll m$. Moreover, the calculation at finite temperature shows that our
model is an excellent approximation in the regime of $\frac{kT}{m}<0.05$ where terms proportional
to $\left( kT\right)  ^{6}$ are subdominants \cite{Gies2000}. It is worth mentioning that
the consistency between the two approaches was not obvious a priori since we
use the $1$-loop Heisenberg-Euler effective action at zero temperature.

Let us move on and take into account the $2$-loops Heisenberg-Euler effective action.
According to \cite{Kors}, the effective Lagrangian is%
\[
L^{2}=F+\bar{c}_{1}F^{2}+\bar{c}_{2}G^{2}+\bar{c}_{3}F^{3}+\bar{c}_{4}%
FG^{2}+...
\]
with%
\begin{align*}
\bar{c}_{1}  &  =c_{1}+\frac{\alpha^{3}}{\pi m^{4}}\frac{2^{6}}{3^{4}%
}\text{,\ \ \ }\bar{c}_{2}=c_{2}+\frac{\alpha^{3}}{\pi m^{4}}\frac
{263}{2\times3^{4}}\text{,}\\
\text{ }\bar{c}_{3}  &  =c_{3}-\frac{\alpha^{4}}{m^{8}}\frac{2^{3}%
\times23\times53}{3^{4}\times5^{2}}\text{,\ }\bar{c}_{4}=c_{4}-\frac
{\alpha^{4}}{m^{8}}\frac{2^{5}\times541}{3^{4}\times5^{2}}.
\end{align*}

Performing the same approximations as before we obtain%
\begin{align}
\rho &  =\frac{\pi^{2}}{15}\left(  kT\right)  ^{4}+\left[  \frac{44\alpha
^{2}\pi^{2}}{675}+\frac{391 \pi\alpha^{3}}{5\times
3^{5}}\right]  \frac{\left(  H-F\right)  }{m^{4}}\left(  kT\right)
^{4}\nonumber\\
&  +\frac{2^{6}\times37\alpha^{3}\pi^{3}}{3^{3}\times5^{2}\times7}%
\frac{F\left(  H-F\right)  }{m^{8}}\left(  kT\right)  ^{4}+\mathcal{O}(3),
\label{rhoHES2}%
\end{align}
where the new term arises due the contribution of $2$-loops effective action.
As expected, this new term represents a small correction (about $1\%$)
compared with $\frac{44\alpha^{2}\pi^{2}}{675}$. However, for $\frac{F}{m^{4}%
}<0.05$ its contribution becomes dominant when compared to the last term in
(\ref{rhoHES2}). Restoring the (Gaussian) units we have%
\[
\frac{F}{m^{4}}\rightarrow\frac{\hbar^{3}c^{3}F}{m^{4}c^{8}}\rightarrow
7\times10^{-26}\left(  \frac{E^{2}-B^{2}}{2}\right)  .
\]
Thus,
\[
\frac{F}{m^{4}}<0.05\Rightarrow\left\{
\begin{array}
[c]{c}%
B<1.2\times10^{12}\text{ }G\\
E<3.6\times10^{14}\text{ }V/cm
\end{array}
\right.  .
\]
As almost all electric and magnetic fields known respect the constraints above
(exception for magnetars) the new term in (\ref{rhoHES2}) is dominant over
the last one.

Finally, it is important to emphasize that if there is consistency between our
approach and the effective action at finite temperature, then the thermal three
(or more) loops calculation must derive the correction $\frac{391 \pi\alpha^{3}}{5\times3^{5}}$.
As far as we know this calculation has never been done.

\section{Born-Infeld Lagrangian}

The most well known extension of Maxwell electrodynamics is the
Born-Infeld electrodynamics which is produced by the following Lagrangian
\cite{BI,BI1}:%
\[
L_{BI}=b^{2}\left[  1-\sqrt{1-\frac{2F}{b^{2}}-\frac{G^{2}}{b^{4}}}\right],
\]
where $b^{2}$\ is an arbitrary real parameter with dimension of energy density.
The Maxwell electrodynamics is recovered when $b\rightarrow \infty$.

What makes the Born-Infeld NLED interesting are its special features. For
example, $L_{BI}$ has an upper limit value of the fields avoiding the problem
of divergence of the electrostatic self-energy of a point charge. It is the
only physical NLED which presents no birefringence \cite{CLP}. Besides, a
generalization of Born-Infeld model describes the electrodynamics on D-branes
in the context of open string theories \cite{FraTsey}.

Although Born-Infeld Lagrangian has no restrictions it possesses a characteristic
energy scale associated with the $b$ parameter. This characteristic scale sets
$M=\sqrt{b}$, and thus our approach is valid only in the regime of $kT \ll \sqrt{b}$.

\begin{figure}[th]
\includegraphics[height=6.3cm, width=8.2cm]{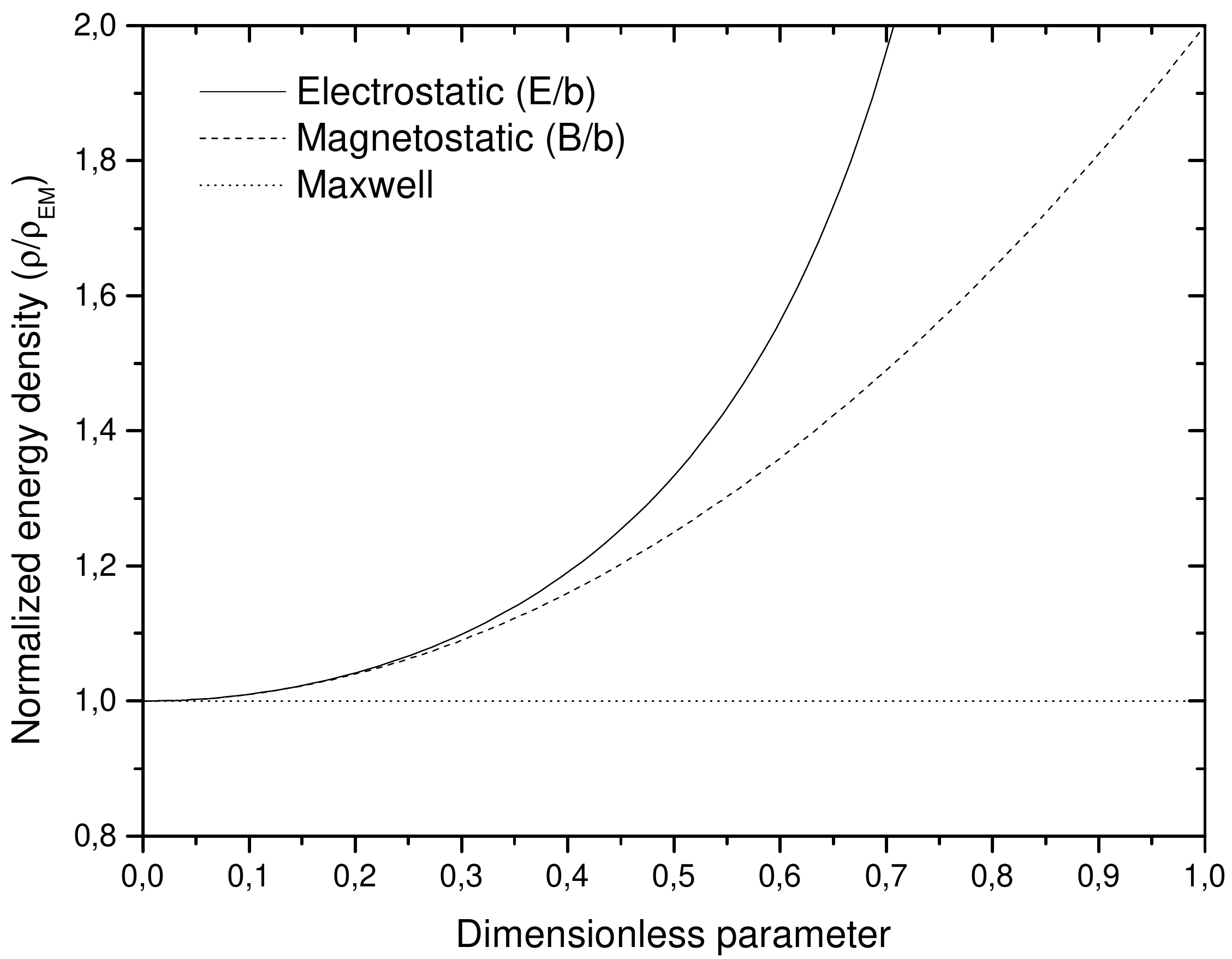}
\caption{Comparison between the energy densities for Born-Infeld and Maxwell
electrodynamics. The energy densities are plotted in terms of
dimensionless parameters $\frac{E_{1}}{b}$ and $\frac{B_{1}}{b}$}%
\label{fig1}%
\end{figure}

For the Born-Infeld Lagrangian, $\delta=0$ and (\ref{z main}) is given by%

\[
z_{+}^{BI}=z_{-}^{BI}=-\frac{1}{2F-b^{2}}.
\]
In order to simplify the manipulation and for insight purposes we will analyze only
when the uniform background field is purely magnetic or electric. In this case, we can
write the energy densities for Born-Infeld model as%
\begin{equation}
\rho_{BI}\left(  E_{1},T\right)  =\frac{\pi^{2}}{15}\left(  \frac{1}%
{1-\frac{E_{1}^{2}}{b^{2}}}\right)  \left(  kT\right)  ^{4} \label{rho3}%
\end{equation}
and%
\begin{equation}
\rho_{BI}\left(  B_{1},T\right)  =\frac{\pi^{2}}{15}\left(  \frac{B_{1}^{2}%
}{b^{2}}+1\right)  \left(  kT\right)  ^{4}. \label{rho1}%
\end{equation}
It is important to look for the validity interval dictated by the constraints
(\ref{Cons}). For the electrostatic case $E_{1}<b$, which is in agreement with
the role that the constant $b$ lays. For the magnetostatic case, the
constraints are always satisfied, which means that $B_{1}$ can assume any
value. In order to compare the behaviors, a dimensionless comparison between the
two $\rho_{BI}$ and the standard $\rho_{EM}$ is formed in figure \ref{fig1}.


Figure \ref{fig1} shows that, for a given $kT$, the two $\rho_{BI}$ increase as the
dimensionless field parameters increase and they are always greater than
$\rho_{EM.}$. Physically, this phenomenon is due to the interaction of the
photons with the background fields leading to an excitation
of the gas i.e., the background fields transfer energy to the photon gas. It is
worth noting that this energy transfer is more effective in the electrostatic case.
Similar situation occurs with effective Heisenberg-Euler Lagrangians since the
heat-bath engenders an extra excitation for the photon gas\footnote{See the
dominant correction in (\ref{rhoHES2}).}.

\section{Final Remarks}

In this paper the properties of a photon gas were studied in the context of a
general NLED neglecting the photon self-interaction. As usual, the
Bose-Einstein statistics was used and the calculation of the partition
function was performed. The grand canonical potential was obtained considering an
uniform background field, and for this case, we verified that although the
energy density and pressure depend on the background field, the equation of
state remains unaltered, namely $\rho=3p$.

Applications of the results were done to Heisenberg-Euler and Born-Infeld
nonlinear electrodynamics. Using the $1$-loop Heisenberg-Euler
effective action at zero temperature we were able to recover the results
obtained in the thermal $2$-loops approach \cite{Gies2000}. Besides, including
corrections at $2$-loops (zero temperature) we predicted a new term which
should be dominant in second order for the photon gas energy density. It would
be interesting, in a future work, to verify if this new term can be derived from
the thermal $3$-loop Heisenberg-Euler effective action. For the Born-Infeld model,
the interaction of the photons with the background field might produce
an excitation of the gas.

The approach developed in this paper can be used to test the linearity of electrodynamics or
set constraints for specific NLED. For example, one may look for
deviations of Planck spectrum and/or modifications of energy density in situations
of strong electromagnetic field (e.g. around magnetars). Although our approach
applies to any NLED it does not take into account the self interaction of
photons. This approximation limits the scope of possible applications. A
solution is to choose a specific NLED and apply the procedure of quantization
at finite temperature to obtain a complete description of the black body
phenomenon \cite{ProPod2}. In this situation, it would be possible, for
example, to analyze the effects of a NLED in primordial cosmology.

\section*{{\protect\small Acknowledgements}}

The authors acknowledge I. Bialynicki-Birula, H. Gies and R. R. Cuzinatto
for their useful comments. P. Niau Akmansoy is grateful to CAPES-Brazil for financial support.
L. G. Medeiros acknowledges FAPERN-Brazil for financial support.


\appendix

\section{Birefringence Phenomenon}

\label{apen}

Following the procedure described in \cite{Birula}, the amplitude of
electromagnetic field for a photon is given by
\[
\varepsilon_{\mu\nu}=k_{\mu}\varepsilon_{\nu}-k_{\nu}\varepsilon_{\mu}%
\]
where%
\[
\varepsilon_{\mu}=\alpha a_{\mu}+\beta\hat{a}_{\mu}+\gamma k_{\mu}.
\]
with $a^{\mu}\equiv F^{\mu\nu}k_{\nu}$ and $\hat{a}^{\mu}\equiv\hat{F}^{\mu
\nu}k_{\nu}$. Using the relations $k_{\mu}=\left(  \omega,-\vec{p}\right)  $,
$a_{\mu}=\left(  a_{0},\vec{a}\right)  $ and $\hat{a}_{\mu}=\left(  \hat
{a}_{0},\overrightarrow{\hat{a}}\right)  $, we can write the electric field
$E_{i}=\varepsilon_{i0}$ as%
\begin{equation}
\vec{E}=\alpha\vec{v}+\beta\vec{u} \label{Ele photon}%
\end{equation}
where $\vec{v}=-\left(  a_{0}\vec{p}+\omega\vec{a}\right)  $ and $\vec
{u}=-\left(  \hat{a}_{0}\vec{p}+\omega\overrightarrow{\hat{a}}\right)  $ are
LI vectors. The equation (\ref{Ele photon}) states that the electric field for
a photon is always located on the plane generated by $\vec{v}$ and $\vec{u}$.

On the other hand, the two dispersion relations $\omega_{\pm}$ are obtained
imposing non-trivial solution to the system
\begin{equation}
\left(
\begin{array}
[c]{cc}%
Mk^{2}-L_{FF}a^{2} & GL_{FF}k^{2}-NL_{FG}\\
GL_{GG}k^{2}-L_{FG}a^{2} & Mk^{2}-NL_{GG}%
\end{array}
\right)
\begin{pmatrix}
\alpha\\
\beta
\end{pmatrix}
=%
\begin{pmatrix}
0\\
0
\end{pmatrix}
, \label{System M}%
\end{equation}
with
\begin{align*}
M  &  \equiv L_{F}+GL_{FG}\text{, \ }N\equiv a^{2}+2Fk^{2}\text{\ }\\
k^{2}  &  \equiv k_{\mu}k^{\mu}\text{ \ \ and \ }a^{2}\equiv a_{\mu}a^{\mu}.
\end{align*}
Each dispersion relations, $\omega_{+}$ and $\omega_{-}$, is associated with a
respective vector $\vec{V}_{+}=\left(  \alpha_{+},\beta_{+}\right)  $ and
$\vec{V}_{-}=\left(  \alpha_{-},\beta_{-}\right)  $ which, in the language of
eigenvalue problems, are eigenvectors.\ These two vectors define the LI set%
\[
\vec{E}_{+}=\alpha_{+}\vec{v}+\beta_{+}\vec{u}\text{ \ \ and \ \ }\vec{E}%
_{-}=\alpha_{-}\vec{v}+\beta_{-}\vec{u}\text{ .}%
\]

Thus, any photon whose electric field $\vec{E}$ is in the direction $\vec
{E}_{\pm}$ propagates with a dispersion relation $\omega_{\pm}$. Besides, the
field $\vec{E}$ of any photon can be decomposed in%
\[
\vec{E}=\bar{\alpha}\vec{E}_{+}+\bar{\beta}\vec{E}_{-},%
\]
producing the phenomenon of birefringence.

\end{document}